\documentclass[10pt,a4paper]{article}
\usepackage{amsmath}
\usepackage{amssymb}
\begin{document}
\newcommand{\om}{{\omega}}
\newcommand{\Ai}{{\rm{Ai}}}
\newcommand{\Tr}{{\rm{Tr}}}
\newcommand{\+}{\!+\!}
\newcommand{\m}{\!-\!}
\newcommand{\half}{{\lower0.25ex\hbox{\raise.6ex\hbox{\the\scriptfont0 1}\kern-.2em\slash\kern-.1em\lower.25ex\hbox{\the\scriptfont0 2}}}}
\newcommand{\quar}{{\lower0.25ex\hbox{\raise.6ex\hbox{\the\scriptfont0 1}\kern-.2em\slash\kern-.1em\lower.25ex\hbox{\the\scriptfont0 4}}}}
\newcommand{\eighth}{{\lower0.25ex\hbox{\raise.6ex\hbox{\the\scriptfont0 1}\kern-.2em\slash\kern-.1em\lower.25ex\hbox{\the\scriptfont0 8}}}}
\newcommand{\othird}{{\lower0.25ex\hbox{\raise.6ex\hbox{\the\scriptfont0 1}\kern-.2em/\kern-.1em\lower.25ex\hbox{\the\scriptfont0 3}}}}
\begin{center}
{\LARGE\bf 
Integrability, Random Matrices and Painlev\'e Transcendents}
\end{center}
\vskip2.0cm
\begin{center}
{N.S. Witte,\footnote{\large E-mail: {\tt nsw@ms.unimelb.edu.au}}\\
{\it Department of Mathematics and Statistics}\\
{\it \& School of Physics, University of Melbourne,}\\
{\it Parkville, Victoria 3052, AUSTRALIA.}}\\
\vskip0.3cm
{P.J. Forrester},\\
{\it Department of Mathematics and Statistics}\\
{\it University of Melbourne}\\
{\it Parkville, Victoria 3052, AUSTRALIA}\\
\vskip0.3cm
and {Christopher M. Cosgrove}\\
{\it School of Mathematics and Statistics}\\
{\it University of Sydney}\\
{\it Sydney, NSW 2006, AUSTRALIA.}
\end{center}
\bigskip
\begin{abstract}
The probability that an interval $I$ is free of eigenvalues in a matrix
ensemble with unitary symmetry is given by a Fredholm determinant. When
the weight function in the matrix ensemble is a classical weight function,
and the interval $I$ includes an endpoint of the support, Tracy and Widom
have given a formalism which gives coupled differential equations for
the required probability and some auxilary quantities. We summarize and
extend earlier work by expressing the probability and some of the
auxilary quantities in terms of Painlev\'e transcendents.
\end{abstract}

\medskip
\begin{center}
Dedicated to Martin Kruskal on the occasion of his $75$th Birthday.
\end{center}
\bigskip
\section{Motivations from Physics and Mathematics}

Diverse branches of physics
and mathematics provide many examples of quantities which are known to have
the same statistical properties as the eigenvalues of large random matrices.
This is the case for quantum chaos --- the study of the quantum mechanics of
systems which exhibit chaos at the classical level. Here random matrices
have emerged as a paradigm, in that those systems for which the underlying
classical mechanics is chaotic are observed to have energy levels with the
same statistical properties as the eigenvalues of large random matrices,
while systems with integrable classical dynamics do not. Many examples are
known from the study of two-dimensional billiard systems --- a single particle
in a closed domain in two dimensions. One of the most celebrated is the Sinai
billiard, which is a square billiard with a hard wall circular scatterer in the
centre. Numerical studies show that this system, which is known to be
chaotic at the classical level, has energy levels with the same statistical
properties of large real symmetric matrices (see e.g.~\cite{GM-GW-99}). 
The status of our understanding is reflected in a conjecture by
Bohigas-Giannoni-Schmit which states: 
\begin{quote}
 ``The spectra of time-reversal invariant systems whose classical analogues
  are K-systems show the same fluctuation properties as predicted by the
  GOE.''
\end{quote}
Here GOE refers to the Gaussian orthogonal ensemble of random real symmetric
matrices and K-systems are the most strongly mixing classical systems. The
time reversal symmetry is responsible for the requirement that the random
matrices have real entries.

Another fascinating and enigmatic connection is that of random matrix theory 
and number theory, which concerns the statistics of the large zeros of 
the Riemann zeta function on the critical line. Incredible numerical evidence
of Odlyzko \cite{O-89}
based on the calculation of the critical zeros well into the asymptotic
regime (zero number $10^{20}$ and $10^6$ of its neighbours)
has convincingly shown their statistics to accurately follow 
the Gaussian unitary ensemble (GUE) of random Hermitian matrices predictions.
This connection was first observed in analytic work of Montgomery 
\cite{Mo-73} on the pair 
correlations of the zeros. Denoting 
\begin{equation}
  s_n = \half+iE_n \ ,\qquad
  d(E) = \sum_{n \ge 1} \delta(E-E_n) \ ,
\label{zeta_zero}
\end{equation}
for the zeros and their density respectively,
the asymptotic pair correlation was  evaluated as
\begin{align}
  1+S(\tilde{\epsilon}) \equiv
  \lim_{E \to \infty} {\langle d(E)d(E+\epsilon)\rangle \over 
                       \langle d(E)\rangle \langle d(E+\epsilon)\rangle} 
  & = \delta(\tilde{\epsilon})+1-{\sin^2 \pi\tilde{\epsilon} \over
                                  (\pi\tilde{\epsilon})^2 } \\
  & = 1+S_{GUE}(\tilde{\epsilon})
\label{GUE_hypoth}
\end{align}
where $\tilde{\epsilon} \equiv \epsilon/ \langle d(E) \rangle $. This prediction, 
and other statistical measures of the asymptotic zeros such as spacing
statistics, were tested in the numerical work of Odlyzko with almost
perfect agreement being found at a graphical level.
\section{Matrix Ensembles, Statistical Mechanics of Levels}
We take as an example of a random matrix ensemble the GUE, which in application 
to chaotic quantum systems applies in the case of no time-reversal symmetry 
$ [H,T] \neq 0 $. One constructs a $ N\times N $ Hermitian matrix $ X $ by
taking real diagonal elements $ x_{jj} $ to be independently chosen with p.d.f.
\begin{equation}
   {1\over \sqrt{\pi}} e^{-x^2_{jj}} \ ,
\label{GUE_diag}
\end{equation}
and upper triangular elements $ x_{jk} \equiv u_{jk}+iv_{jk} $
are independently chosen with p.d.f.
\begin{equation}
   {2\over \pi} e^{-2(u^2_{jk}+v^2_{jk})}
   = {2\over \pi} e^{-2|x_{jk}|^2} \ .
\label{GUE_off_diag}
\end{equation}
The joint p.d.f.~for elements of $ X $ is then
\begin{equation}
   P(X)
   \propto \prod^N_{j=1} e^{-x^2_{jj}}
       \prod_{1 \leq j < k \leq N} e^{-2|x_{jk}|^2}
  = \prod_{1 \leq j,k \leq N} e^{-|x_{jk}|^2}
  = e^{-\Tr (X^2)} \ .
\label{pdf_form}
\end{equation}
This distribution function is invariant under arbitrary 
unitary transformations
\begin{equation}
   P(U^{-1}XU) = P(X) \ ,
\label{pdf_inv}
\end{equation}
and hence the naming of the ensemble.
The Gaussian form of the p.d.f.~can be uniquely characterised by the invariance
(\ref{pdf_inv}) and the factorisation property
\begin{equation}
   P(X) = \prod_{j,k} f(x_{jk}) \ .
\label{pdf_factor}
\end{equation}
Alternatively it is the unique p.d.f. which maximises the entropy
\begin{equation}
   S[P] = -\int \mu(dX) P\log P \ ,
\label{pdf_entropy}
\end{equation}
subject to the constraint $ \langle \Tr X^2\rangle = N^2 $.
Primary interest centres on the eigenvalue level statistics and so a 
description in terms of eigenvalues is required. This is achieved by the 
mapping
\begin{equation}
   \half N(N+1) \; \text{elements} \; \{x_{ij}\} \to
   \begin{cases}
   N \; \text{eigenvalues} & \lambda_1 \leq \ldots \leq \lambda_N \\
   \half N(N-1) \; \text{variables} & p_1, \ldots
   \end{cases}
\label{element_map}
\end{equation}
where the $ p_i $ are related to the eigenvectors.
Using the transformation of Hermitian matrices to the diagonal representation
the volume form is correspondingly transformed as
\begin{equation}
    (dX) \equiv \bigwedge^{N}_{j=1} dx_{jj}
            \bigwedge_{1 \leq j < k \leq N} du_{jk}\wedge dv_{jk}
     = (U^{\dagger}dU) \prod_{1 \leq j < k \leq N}(\lambda_j-\lambda_k)^2
                       \bigwedge^{N}_{j=1} d\lambda_j \ .
\label{volume_xfm}
\end{equation}
This eigenvalue and eigenvector dependence thus factorises.
The eigenvalue pdf can then be written as
\begin{equation}
   P(\lambda_1, \ldots,\lambda_N)  = 
   {1\over C_N} e^{-\sum_{j} \lambda^2_j}
   \prod_{1 \leq j < k \leq N}(\lambda_j-\lambda_k)^2 \ .
\label{pdf_eigen}
\end{equation}
This p.d.f. is proportional to the Boltzmann weight factor 
\begin{equation}
   p(x_1,\ldots,x_N) = {1\over C_{\beta N}} \prod^{N}_{j=1} w(x_j)
                       \prod_{1 \leq j < k \leq N} |x_j-x_k|^{\beta}
   = {1\over Z_N} e^{-\beta U(x_1,\ldots,x_N)} \ ,
\label{pdf_boltzmann}
\end{equation}
from equilibrium
statistical mechanics
for a system of $ N $ charges on the line (or circle) at temperature $ 1/\beta $
located at positions $ x_j $, and with the total potential energy
\begin{equation}              
  U(x_1,\ldots,x_N) = -{1\over \beta}\sum_{1 \leq j \leq N} \ln w(x_j)
                      -\sum_{1 \leq j < k \leq N} \ln |x_j-x_k| \ ,
\label{pdf_pot}
\end{equation}
(with $ -\log w(x) = x^2 $)
composed of a 1-body confining potential (the first term) and a 2-body 
electrostatic interaction (the second term).
Consequently a configuration integral can be defined
\begin{equation}
  Z_N = \int^{\infty}_{-\infty} dx_1 \ldots \int^{\infty}_{-\infty}
         dx_N e^{-\beta U} \ .
\label{pdf_Zpf}
\end{equation}

The gap probability, or probability for the exclusion of eigenvalues from an
interval $ I $, is a fundamental statistic underlying the spacing distribution
and is defined by
\begin{equation}
   E_{\beta}(0;I) =
   {1\over Z_N} \int_{\bar{I}} dx_1 \ldots \int_{\bar{I}} dx_N e^{-\beta U} \ ,
\label{gap_prob}
\end{equation}
with $ \bar{I} = (-\infty,\infty)\backslash I $,
however this form is often not suitable to work with. Instead we seek a form
involving the $n$-particle distribution function
\begin{align}
   \rho_{n} (x_1,\ldots,x_n)
 & \equiv \Big< \sum_{1 \leq j_1 \neq \cdots \neq j_n \leq N}
         \delta(x_1-x_{j_1}) \cdots \delta(x_n-x_{j_n}) \Big> \ ,
 \nonumber \\
 & = {N!\over (N\!-\!n)!}{1\over Z_N}
     \int^{\infty}_{-\infty}dx_{n+1} \ldots \int^{\infty}_{-\infty} dx_N\,
     e^{-\beta U(x_1, \ldots ,x_N)} \ .
\label{n-particle}
\end{align}
\noindent
This can be done by introducing the generating functional
\begin{equation}
 Z_N[a] \equiv {1\over Z_N}
   \int^{\infty}_{-\infty}dx_{1} \ldots \int^{\infty}_{-\infty}dx_N
   \prod^{N}_{l=1} \left[1+a(x_l)\right] e^{-\beta U(x_1, \ldots ,x_N)} \ ,
\label{gen_func}
\end{equation}
expanding out the product and using the definition (\ref{n-particle}) to
obtain the formula
\begin{equation}
 Z_N[a] = 1 + \sum^{\infty}_{n=1} {1\over n!}
   \int^{\infty}_{-\infty}dx_{1} \ldots \int^{\infty}_{-\infty}dx_{n}\,
   \prod^{n}_{l=1}a(x_l) \,\rho_{n} (x_1,\ldots,x_n) \ .
\label{gen_func_exp}
\end{equation}
Choosing $ -a(x) $ to be the characteristic function of $ I $ so that
\begin{align}
   a(x) =
   \begin{cases}
    -1 & \text{if}\quad x \in I \\
     0 & \text{if}\quad x \notin I \ ,
   \end{cases}
\label{char_fun}
\end{align}
we see that (\ref{gen_func}) coincides with the definition (\ref{gap_prob})
of $ E_{\beta}(0;I) $ and (\ref{gen_func_exp}) then gives
\begin{equation}
   E_{\beta}(0;I) = 1 + 
   \sum^{\infty}_{n=1} {(-1)^n\over n!}
   \int_{I}dx_{1} \ldots \int_{I}dx_{n}\,
        \rho_{n} (x_1,\ldots,x_n) \ .
\label{gap_exp}
\end{equation}
The probability $E_{\beta}(0;I)$ determines
the p.d.f.~that given there is a particle at $ a_1 $, the nearest neighbour to 
the right is at $ a_2 $, through the formula
\begin{align}
   p_{\beta}(0;(a_1,a_2)) 
   & \equiv {N(N\!-\!1)\over \rho(a_1)Z_N}
     \int_{\bar{I}}dx_{3} \ldots \int_{\bar{I}}dx_{N}\,
          e^{-\beta U(a_1,a_2,x_3, \ldots ,x_N)} \ ,\\
   & = -{1\over \rho(a_1)} {\partial^2\over \partial a_1\partial a_2}
        E_{\beta}(0;(a_1,a_2)) \ .
\label{spacing_pdf}
\end{align}
The distribution $p_{\beta}(0;(a_1,a_2))$ is readily measured empirically,
especially in the bulk region of the spectrum when translation invariance
gives $p_{\beta}(0;(a_1,a_2)) = p_\beta(0;|a_1-a_2|)$.

\section{Fredholm Integral Operators}
For $\beta = 2$ the probability $E_\beta(0;I)$ can be written as the
determinant of a Fredholm integral operator \cite{M-91}. To show this
we make use of the fact that the general $n$-particle distribution
has the determinant representation
\begin{equation}
   \rho_{n} (x_1,\ldots,x_n) =
   \det[K(x_i,x_j)]^{n}_{i,j=1} \ ,
\label{n-part_det}
\end{equation}
where, with $\{p_n(x)\}_{n=0,1,\dots}$ the orthonormal polynomials with respect
to the weight function $w(x)$,
\begin{equation}
  K(x,y)
  \equiv [w(x)w(y)]^{1/2} \sum^{N-1}_{n=0}p_n(x)p_n(y) \ ,
\label{kernel_ops}
\end{equation}
With $a_n$ denoting the coefficient of $x^n$ in $p_n(x)$ 
the Christoffel-Darboux summation gives
\begin{align}
  K(x,y) 
  & = {a_{N-1}\over a_N} [w(x)w(y)]^{1/2} 
      {p_N(x)p_{N-1}(y)-p_{N-1}(x)p_N(y) \over x-y} \ ,
  \nonumber \\
  & = { \phi(x)\psi(y) - \phi(y)\psi(x) \over x-y } 
\label{kernel_CDsum}
\end{align}
The determinant formula (\ref{n-part_det}) substituted in (\ref{gap_exp})
gives a well known \cite{WW-65} expansion formula for the determinant of
an integral operator in terms of its kernel. Explicitly
\begin{equation}
    E(0;I)
     = \det( \Bbb{I}-\Bbb{K} ) 
\label{gap_fredholm}
\end{equation}
where $\Bbb{K}$ is the integral operator on $I$ with kernel $K(x,y)$.

Tracy and Widom \cite{TW-94} have shown that by introducing a number of
auxiliary quantities related to $\Bbb{K}$, for some special choices of
$w(x)$ and $I$ the Fredholm determinant (\ref{gap_fredholm}) can be
specified in terms of the solution of a nonlinear equation. The
auxiliary quantities required are the operators
\begin{equation}
    {\Bbb{K} \over \Bbb{I}-\Bbb{K}} \doteq R(x,y) \ ,
 \qquad 
              {\Bbb{I} \over \Bbb{I}-\Bbb{K}} \doteq \rho(x,y) \  , 
\label{resolv_op}
\end{equation}
where the notation $\Bbb{A} \doteq A(x,y)$ denotes that the integral
operator $\Bbb{A}$ has kernel $A(x,y)$; the pair of functions
\begin{equation}
\begin{split}
   Q(x) & = \int_{I} dy\; \rho(x,y)\phi(y) \doteq (\Bbb{I}-\Bbb{K})^{-1}\phi \ ,
        \\
   P(x) & = \int_{I} dy\; \rho(x,y)\psi(y) \doteq (\Bbb{I}-\Bbb{K})^{-1}\psi \ ,
\end{split}
\label{QP_defn}
\end{equation}
\begin{equation}
   q_j \equiv \lim_{x \to a_j}Q(x) \ , \qquad
   p_j \equiv \lim_{x \to a_j}P(x) \ ;
\label{qp_defn}
\end{equation}
and the scalar products
\begin{align}
     u & = \langle \phi | Q \rangle = \int_{I}dy\; Q(y)\phi(y) \ ,
       \nonumber \\ 
     v & = \langle \psi | Q \rangle = \int_{I}dy\; Q(y)\psi(y)
         = \langle \phi | P \rangle = \int_{I}dy\; P(y)\phi(y) \ ,
       \\
     w & = \langle \psi | P \rangle = \int_{I}dy\; P(y)\psi(y) \ .
       \nonumber
\label{uvw_defn}
\end{align}

The essence of the Tracy \& Widom machinery in this setting is to set up a 
system of partial differential equations for the above quantities with
respect to the endpoints of the interval, defined by $ I=(a_1,a_2) $. 
One set of equations are universal while a second set of differential equations 
describes the coincident functions and their derivatives, and includes specific 
features of the particular orthogonal polynomial system at hand.
In particular one requires the
differential-recurrence relations
\begin{equation}
\begin{split}
  m(x)\phi'(x) & =  A(x)\phi(x)+B(x)\psi(x) \ ,
  \\
  m(x)\psi'(x) & = -C(x)\phi(x)-A(x)\psi(x) \ ,
\end{split}
\label{diff_reccur}
\end{equation}
where $ m,A,B,C $ are polynomials in $ x $ parameterised by
\begin{equation}
\begin{split}
  m(x) & = \mu_{0}+\mu_{1}x+\mu_{2}x^2 \ ,
  \\
  A(x) & = \alpha_{0}+\alpha_{1}x \ ,
  \\
  B(x) & = \beta_{0}+\beta_{1}x  \ ,
  \\
  C(x) & = \gamma_{0}+\gamma_{1}x \ ,
\end{split}
\label{DR_param}
\end{equation}
which is a common structure for
 the classical orthogonal polynomials. 
\section{\mbox{Spacing p.d.f. and Painlev\'e} transcendents}
The classical weights are
\begin{equation}\label{clw}
w(x) = \left\{ \begin{array}{ll}
		e^{-x^2}, & {\rm Gaussian} \\
		x^a e^{-x}, \: \: x >0, & {\rm Laguerre} \\
		(1-x)^a(1+x)^b, \: \: -1 < x < 1, & {\rm Jacobi}, 
		\end{array} \right.
\end{equation}
which together with their corresponding orthonormal polynomials give
rise to functions $\phi(x)$, $\psi(x)$ satisfying (\ref{diff_reccur})
and (\ref{DR_param}). For general parameters in all
these cases, it has
only been possible to characterise $E_2(0;I)$ in terms of a nonlinear
equation when the interval $I$ includes an endpoint of the support
of $w(x)$ and so depends only on one variable. It turns
out that then the auxiliary quantities $q'/q$, $p'/p$
and $R \equiv R(s,s)$, as well as the probability $E_2(0;I)$ can be
expressed in terms of Painlev\'e transcendents.

In Tables 1--3 we list the evaluations for the three classical weights
respectively. Also listed are the specific coefficients (\ref{DR_param}),
and an equation or equations (integrals of motion) relating the
quantities.

To illustrate the construction of these tables, consider the Jacobi weight
with $I = (-1,s)$. Direct application of the Tracy and Widom theory gives
the coupled differential equations
\begin{align}
  [\ln E_{2}]'
  & = -R \ ,
  \label{jacobi-sde:a}\\
  (1\m s^2)q'
  & =   [\alpha_0+\alpha_1 s+v] q
      + [\beta_0+u(2\alpha_1\m 1)] p \ ,
  \label{jacobi-sde:b}\\
  (1\m s^2)p'
  & = - [\gamma_0-w(2\alpha_1\+ 1)] q
      - [\alpha_0+\alpha_1 s+v] p \ ,
  \label{jacobi-sde:c}\\
  u'
  & = q^2 \ ,
  \label{jacobi-sde:d}\\
  v'
  & = qp \ ,
  \label{jacobi-sde:e}\\
  w'
  & = p^2 \ ,
  \label{jacobi-sde:f}\\
  (1\m s^2)R
  & =  [\gamma_0-w(2\alpha_1\+ 1)] q^2
     + [\beta_0+u(2\alpha_1\m 1)]  p^2
     + [\alpha_0+\alpha_1 s+v] 2qp
  \label{jacobi-sde:g}\\
  \left[(1\m s^2)R\right]'
  & =  2\alpha_1 qp \ ,
  \label{jacobi-sde:h}
\end{align}
From this system of differential equations two integrals can be constructed
\cite{WF-00}, which are given in Table \ref{Jacobi-table}, and using these
a sequence of eliminations leads to a second degree second order differential
equation for $ \sigma(s) \equiv (1\m s^2)R(s) $. This was first found in an
indirect manner in \cite{HS-99}. In Table \ref{Jacobi-table} we identify the
solution of this equation with a number of P-VI transcendents, and give that
specific mapping in the first case listed. For this purpose we use 
theory from \cite{CS-93}.
\section{Conclusions}
In this work we have considered the evaluation of the gap probability
$E_2(0;I)$ for the classical weights (\ref{clw}) with $I$ including an
endpoint of the support. This quantity, and the auxiliary quantities
$p'/p$, $q'/q$ and $R \equiv R(s,s)$, are given in terms of
Painlev\'e transcendents thereby extending findings of earlier works
\cite{TW-94,HS-99,WF-00}.
Our approach has been to construct all the integrals of the motion which
allows us to reduce the system of differential equations arising in the 
Tracy and Widom formalism to a single second order ODE by the most direct 
path possible. Furthermore we have displayed the appearance of the 
Painlev\'e transcendents in the random matrix context in a transparent
manner, which we consider to be essential for an understanding of the relevance
of integrability in this situation.

\bibliographystyle{siam}
\bibliography{moment,random_matrices,nonlinear}
\vfill\eject

\renewcommand{\arraystretch}{1.5}
\begin{table}

\begin{center}
\begin{tabular}{|c|}
\hline
\underline{Gaussian weights and Hermite orthogonal polynomials} \\
 $ w(s) = e^{-s^2} $,
 $ m(s) = 1 $,
 $ A(s) = -1 $,
 $ B(s) = C(s) = \sqrt{2N} $ \\
\hline
\hline
 $ I = (s,\infty) $ \\
\hline
\begin{minipage}[b]{10cm}
\begin{equation*}
   \sqrt{2N}[u-w] + 2uw = qp
\end{equation*}
\end{minipage} \\
\hline
 $ E_2(s) = \exp\left(-\int^{\infty}_{s}dt\, R(t) \right) $ \\
\hline
\begin{minipage}[b]{12cm}{
\begin{equation*}
\begin{split}
   R(s) \underset{s \to \infty}{\sim} & \;
   {2^{N-1} \over \sqrt{\pi}(N\m 1)!} s^{2N-2}e^{-s^2} \\
   {q' \over q} \underset{s \to \infty}{\sim} &\;
   -s + {N \over s} \\
   {p' \over p} \underset{s \to \infty}{\sim} &\;
   -s + {N\m 1 \over s}
\end{split}
\end{equation*}
}\end{minipage} \\
\hline
 $ \om(s) $ PIV
 $ \quad \alpha = 2N-1, \; \beta = 0 $ \\
\hline
\begin{minipage}[b]{10cm}{
\begin{equation*}
\begin{split}
   {q' \over q} &
   = -s-\om-{2N\om \over \half\om'-\half\om^2-s\om}
   \\
   {p' \over p} &
   = -\half \om+{1\over 2\om}\om'
   \\
   R & 
   = -\half(s^2\m 2N)\om-\half s\om^2-\eighth \om^3
     +{1\over 8\om}(\om')^2
\end{split}
\end{equation*}
}\end{minipage} \\
\hline
 $ \om(s) $ PIV
 $ \quad \alpha = -N\pm 1, \; \beta = -2N^2 $ \\
\hline
\begin{minipage}[b]{10cm}{
\begin{equation*}
   R = -{N^2\over 2\om}-Ns-\half(s^2\+ N)\om-\half s\om^2-\eighth \om^3
        +{1\over 8\om}(\om')^2
\end{equation*}
}\end{minipage} \\
\hline
\end{tabular}
\end{center}
\bigskip               
\caption{               
\mbox{Painlev\'e transcendents in the Gaussian Unitary Ensemble} 
}        
\label{Hermite-table} 
\bigskip              
\end{table}

\renewcommand{\arraystretch}{1.5}
\begin{table}

\begin{center}
\begin{tabular}{|c|}
\hline
\underline{Laguerre weights and orthogonal polynomials} \\
\begin{minipage}[b]{10cm}{
\begin{equation*}
\begin{split}
  w(s) & = s^ae^{-s} \\
  m(s) & = s \\
  A(s) & = -\half a-N+\half s \\
  B(s) & = C(s) = \sqrt{N(N\+ a)}
\end{split}
\end{equation*}
}\end{minipage} \\
\hline
\hline
 $ I = (0,s)  $ \\
\hline
\begin{minipage}[b]{10cm}
\begin{equation*}
   \sqrt{N(N\+ a)}[w-u] + uw = sqp-sR
\end{equation*}
\end{minipage} \\
\hline
 $ E_2(s) = \exp\left(-\int_{0}^{s}dt\, R(t) \right) $ \\
\hline
\begin{minipage}[b]{12cm}{
\begin{equation*}
\begin{split}
   \sigma(s) \equiv sR(s) \underset{s \to 0}{\sim} & \;
   {\Gamma(N\+ a\+ 1) \over
    (N\m 1)!\Gamma(a\+ 1)\Gamma(a\+ 2)} s^{a+1}e^{-s}
   \left\{ 1 - {2N\m 2 \over a\+2}s \right\} \\
   {q' \over q} \underset{s \to 0}{\sim} &\;
   {a \over 2s} - {2N\m 1\+ a \over 2(a\+ 1)} \\
   {p' \over p} \underset{s \to 0}{\sim} &\;
   {a \over 2s} - {2N\+ 1\+ a \over 2(a\+ 1)}
\end{split}
\end{equation*}
}\end{minipage} \\
\hline
\begin{minipage}[b]{12cm}{
\begin{alignat*}{5}
 \om(s) \; {\rm PV} \;
 & \quad &  \alpha =\quad		& \quad &  \beta =\quad
 & \quad &  \gamma =\quad 		& \quad &  \delta =\quad\\
 &  & \half(1\m a)^2		&  & 0 
 &  & -2N\m a 			&  & -\half \\
 &  & \half  			&  & -\half a^2 
 &  &  2N\+ a     		&  & -\half \\
 &  & \half(1\m a\m N)^2  	&  & -\half N^2 
 &  & -a  			&  & -\half \\
 &  & \half(1\m N)^2 		&  & -\half(N\+ a)^2 
 &  & a 			&  & -\half
\end{alignat*}
}\end{minipage} \\
\hline
\begin{minipage}[b]{12cm}{
\begin{equation*}
\begin{split}
   {q' \over q} &
   = {(a\m 1)\om-a \over 2s}
     -{2N\om \over (\om\+ \om')s+(a\m 1)\om(\om\m 1)}
     +{\om'\m 1 \over 2(\om\m 1)} \\
   {p' \over p} &
   = {(a\m 1)\om-a \over 2s}
     +{2(N\+ a)\om \over (\om\m \om')s-(a\m 1)\om(\om\m 1)}
     +{\om'\+ 1 \over 2(\om\m 1)} \\
   sR & 
   = -{1\over 4\om}\left[{s\om' \over \om\m 1}-\om\right]^2
     +\quar a^2\om+\half (2N\+ a)s{\om \over \om\m 1}
     +\quar s^2{\om \over (\om\m 1)^2} 
\end{split}
\end{equation*}
}\end{minipage} \\
\hline
\end{tabular}
\end{center}
\bigskip               
\caption{               
\mbox{Painlev\'e transcendents in the Laguerre Unitary Ensemble}
}        
\label{Laguerre-table} 
\bigskip              
\end{table}

\renewcommand{\arraystretch}{1.5}
\begin{table}

\begin{center}
\begin{tabular}{|c|}
\hline
\underline{Jacobi weights and orthogonal polynomials} \\
\begin{minipage}[b]{14cm}{
\begin{equation*}
\begin{split}
 w(s) & = (1\m s)^a(1\+ s)^b \; ,\;
 m(s) = 1\m s^2 \; ,\;
 A(s) = {b^2\m a^2 \over 2(2N\+ a\+ b)}
      - {2N\+ a\+ b\over 2}s \equiv \alpha_0+\alpha_1 s \\
 B(s) & = {2\sqrt{N(N\+ a)(N\+ b)(N\+ a\+ b)}
        \over 2N\+ a\+ b}
        \sqrt{ 2N\+ a\+ b\+ 1 \over 2N\+ a\+ b\m 1 }
      \equiv \beta_0 \\
 C(s) & = {2\sqrt{N(N\+ a)(N\+ b)(N\+ a\+ b)}
        \over 2N\+ a\+ b}
        \sqrt{ 2N\+ a\+ b\m 1 \over 2N\+ a\+ b\+ 1 }
      \equiv \gamma_0
\end{split}
\end{equation*}
}\end{minipage} \\
\hline
\hline
 $ I = (-1,s) $ \\
\hline
\begin{minipage}[b]{12cm}{
\begin{equation*}
\begin{split}
   \sigma(s) \equiv (1-s^2)R &
   = -(2N\+ a\+ b)v \\
   [\beta_0+u(2\alpha_1\m 1)][\gamma_0-w(2\alpha_1\+ 1)] &
   = \beta_0\gamma_0-(1\m s^2)\sigma'-s\sigma
     +{\alpha_0 \over \alpha_1}\sigma+{1\over 4\alpha_1^2}\sigma^2
\end{split}
\end{equation*}
}\end{minipage} \\
\hline
 $ E_2(s) = \exp\left(-\int_{-1}^{s}dt\, R(t) \right) $ \\
\hline
\begin{minipage}[b]{15cm}{
\begin{equation*}
\begin{split}
   \sigma(s) \underset{s \to -1}{\sim} & \;
   {\Gamma(N\+ a\+ b\+ 1)\Gamma(N\+ b\+ 1) \over 
    2^b(N\m 1)!\Gamma(N\+ a)\Gamma(b\+ 1)\Gamma(b\+ 2)} (s\+ 1)^{b+1}
   \left\{ 1 - {2N^2\+ 2N(a+b)\+ ab\m b \over 2(b\+2)}(s\+1) \right\} \\
   {q' \over q} \underset{s \to -1}{\sim} &\;
   {b \over 2(s\+ 1)} - {2N^2\+2N(a\+ b\+ 1)\+ a(b\+ 1) \over 4(b\+ 1)} \\
   {p' \over p} \underset{s \to -1}{\sim} &\;
   {b \over 2(s\+ 1)} - {2N^2\+2N(a\+ b\m 1)\+ a(b\m 1)\m 2b \over 4(b\+ 1)}
\end{split}
\end{equation*}
}\end{minipage} \\
\hline
\begin{minipage}[b]{13cm}{
\begin{alignat*}{5}
 & \quad &   \om(x) \; {\rm PVI} \quad \alpha =\quad
 & \quad &  \beta =\quad
 & \quad &  \gamma =\quad		& \quad &  \delta =\quad\\
 &  & \half 				&  & -\half a^2 
 &  & \half b^2 			&  & \half[1-(2N\+ a\+ b)^2] \\
 &  & \half[1-(2N\+ a\+ b)]^2     	&  & -\half b^2 
 &  & \half a^2 			&  & \half \\
 &  & \half[1\m a]^2     		&  & 0 
 &  & \half(2N\+ a\+ b)^2 		&  & \half[1-b^2] \\
 &  & \half[1\m b]^2     		&  & -\half(2N\+ a\+ b)^2  
 &  & 0  				&  & \half[1-a^2] \\
 &  & \half[1\m N\m a\m b]^2     	&  & -\half(N\+ b)^2  
 &  & \half(N\+ a)^2  			&  & \half[1-N^2] \\
 &  & \half[1\m N]^2     		&  & -\half(N\+ a)^2  
 &  & \half(N\+ b)^2  			&  & \half[1-(N\+ a\+ b)^2] \\
 &  & \half[1\m N\m a]^2     		&  & -\half N^2  
 &  & \half(N\+ a\+ b)^2  		&  & \half[1-(N\+ b)^2] \\
 &  & \half[1\m N\m b]^2     		&  & -\half (N\+ a\+ b)^2  
 &  & \half N^2  			&  & \half[1-(N\+ a)^2]
\end{alignat*}
}\end{minipage} \\
\hline
\begin{minipage}[b]{15cm}{
\begin{equation*}
\begin{split}
   s & = 2x-1 \\
   (1\m s^2){q' \over q} &
   = x\m 1 + \om +(2N\+ 1\+ a\+ b){x(x\m 1) \over \om\m x}
     + x(1\m x){\om' \over \om\m x}
   \\
   (1\m s^2){p' \over p} &
   = x\m 1 + \om -(2N\m 1\+ a\+ b){x(x\m 1) \over \om\m x}
     + x(1\m x){\om' \over \om\m x}
   \\
   \sigma & 
   = {x^2(x\m 1)^2 \over 2\om(\om\m 1)(\om\m x)}
     \left[\om'-{\om(\om\m 1) \over x(x\m 1)}\right]^2
            -\half a^2{x \over \om}+\half b^2{x\m 1 \over \om\m 1}
            +\half(2N\+ a\+ b)^2{x(1\m x) \over \om\m x}
\end{split}
\end{equation*}
}\end{minipage} \\
\hline
\end{tabular}
\end{center}
\bigskip               
\caption{               
\mbox{Painlev\'e transcendents in the Jacobi Unitary Ensemble}
}        
\label{Jacobi-table} 
\bigskip              
\end{table}

\end{document}